\documentclass[secnumarabic, twocolumn, floatfix, aps, superscriptaddress, prb]{revtex4-1}

\usepackage{graphicx}

\begin{document}

\title{Transport of dipolar excitons in (Al,Ga)N/GaN quantum wells}
\author{F. Fedichkin}
\affiliation{Laboratoire Charles Coulomb, UMR 5221
 CNRS-Universit\'{e} de Montpellier, 34095 Montpellier , France}
\author{P. Andreakou}
\affiliation{Laboratoire Charles Coulomb, UMR 5221
 CNRS-Universit\'{e} de Montpellier, 34095 Montpellier , France}
\author{B. Jouault}
\affiliation{Laboratoire Charles Coulomb, UMR 5221
 CNRS-Universit\'{e} de Montpellier, 34095 Montpellier , France}
\author{M. Vladimirova}
\affiliation{Laboratoire Charles Coulomb, UMR 5221
 CNRS-Universit\'{e} de Montpellier, 34095 Montpellier , France}
\author{T. Guillet}
\affiliation{Laboratoire Charles Coulomb, UMR 5221
 CNRS-Universit\'{e} de Montpellier, 34095 Montpellier , France}
 \author{C. Brimont}
\affiliation{Laboratoire Charles Coulomb, UMR 5221
 CNRS-Universit\'{e} de Montpellier, 34095 Montpellier , France}
\author{P. Valvin}
\affiliation{Laboratoire Charles Coulomb, UMR 5221
 CNRS-Universit\'{e} de Montpellier, 34095 Montpellier , France}
 \author{T. Bretagnon}
\affiliation{Laboratoire Charles Coulomb, UMR 5221
 CNRS-Universit\'{e} de Montpellier, 34095 Montpellier , France}
\author{A. Dussaigne}
\affiliation{CRHEA-CNRS, Rue Bernard Gregory, 06560 Valbonne, France}
\author{N. Grandjean}
\affiliation{Institute of Condensed Matter Physics, EPFL, CH-1015 Lausanne, Switzerland}
\affiliation{CRHEA-CNRS, Rue Bernard Gregory, 06560 Valbonne, France}
\author{P. Lefebvre}
\affiliation{Laboratoire Charles Coulomb, UMR 5221
 CNRS-Universit\'{e} de Montpellier, 34095 Montpellier , France}
\begin{abstract}
We investigate the transport of dipolar indirect excitons along the growth plane of polar
(Al,Ga)N/GaN quantum well structures by means of spatially- and
time-resolved photoluminescence spectroscopy.
The transport in these strongly disordered quantum wells is activated by dipole-dipole repulsion. The latter induces an emission blue shift that increases linearly with exciton density, whereas the radiative recombination rate increases exponentially.
Under continuous, localized excitation, we measure a continuous red shift of the emission, as excitons propagate away from the excitation spot. This shift corresponds to a steady-state gradient of exciton density, measured over several tens of micrometers.
Time-resolved micro-photoluminescence experiments provide information on the dynamics of recombination and transport of dipolar excitons. We account for the ensemble of experimental results by solving the nonlinear drift-diffusion equation. Quantitative analysis suggests that in such structures,  exciton  propagation on the scale of 10 to 20 microns is mainly driven by diffusion, rather than by drift, due to the strong disorder and the presence of nonradiative defects.
Secondary exciton creation, most probably by the intense higher-energy luminescence, guided along the sample plane, is shown to contribute to the exciton emission pattern on the scale up to 100 microns.
The exciton propagation length is strongly temperature dependent, the emission being quenched beyond a critical distance governed by nonradiative recombination.

\end{abstract}

\maketitle



\section{Introduction}
\label{sec:intro}
In the last years, indirect excitons (IXs) have attracted considerable
attention because they constitute a model-system for studies of cold  gases of dipolar bosons
 in a solid-state environment \cite{snoke}.
An IX is composed of an electron and a
hole, bound by Coulomb attraction but engineered in such a way that
 that a
potential barrier separates them along one direction in space.
%
%
One of the most interesting and most studied realizations of IXs has
been obtained in  (Al,Ga)As/GaAs coupled quantum wells (CQWs) under
externally applied electric field \cite{Butov1999PRB} .
In such structures the electric field applied along the structure
axis tilts the conduction and valence band edges, so that the
optical recombination takes place between an electron  and a hole
confined in different quantum wells (QWs), which form an IX.
For typical electric fields of several volts per micrometer, red
shift (quantum confined Stark effect) reaches tens of
meV.\cite{Voros2005PRL,Butov1999PRB}
The separation between electron and hole within an IX increases
drastically its radiative lifetime.
Radiative lifetimes as large as tens of microseconds have been reported
in the literature. \cite{Voros2005PRL,Butov1999PRB}
%
%
It has been shown, that if the thermalization time of IXs is shorter than their lifetime,
the IXs can cool down to temperatures comparable or smaller than the degeneracy
temperature, thus allowing for condensation into a coherent
state.\cite{ButovStimScat, KeldyshKozlov, Butov1998,HighNature,DubinEvidencearXiv:1304.4101v1}

Another important feature of IXs, is their intrinsic
dipole moment, aligned perpendicular to the QW plane.
These dipoles constitute an ideal driving force for
putting IXs in motion, since an IX cloud rapidly
expands along the QW plane under its own repulsive
forces.\cite{Rapaport2006PRB, Voros2005PRL,Butov2004}
Dipole-dipole interaction also leads
to the screening of the electric field
and induces a density-dependent blue shift of the exciton energy \cite{Zimmermann}, at least below the Mott density above which excitonic effects are washed out.\cite{Kappei}
Blue shifts of IX energy up to several meVs and propagation
lengths reaching a hundred of microns have been observed in
 GaAs/(Al,Ga)As CQWs at low temperatures.\cite{Voros2005PRL,Alloing}

In-plane motion of IXs can also be controlled by external voltages.
This can be used to collect and manipulate IXs in
traps,\cite{Huber,Chen2006,High2009PRL} lattices~\cite{Remeika2012APL}
and narrow channels\cite{Cohen2011PRL} which are similar
to the traps and lattices used in the studies of cold atoms.\cite{Cornell,Ketterle}
This degree of control opens new possibilities for studying
fundamental physical properties of IXs.\cite{Rontani2009PRB}
The efficient control of IXs by external voltages
also paves the way to novel optoelectronic devices based on
excitonic properties, as recently
demonstrated.~\cite{High2007OptLett,Winbow2007,Grosso2009,Kuznetsova2010OptLett, Andreakou2014APL}
However, the development of such devices
operating at room temperature seems difficult, at least with
GaAs/(Al,Ga)As heterostructures. This is mainly due to the relatively small exciton
binding energy.
Indeed,exciton binding energy is equal to $4$ meV in bulk GaAs,
which corresponds to a temperature of $45$~K. \cite{Fu1991PRB}
It is enhanced by quantum confinement in QWs but it is still reduced
down to a few meV in CQWs, because of the on-axis separation of the
electron-hole pair.

\begin{figure}[ht!]
\center{\includegraphics[width=1\linewidth]{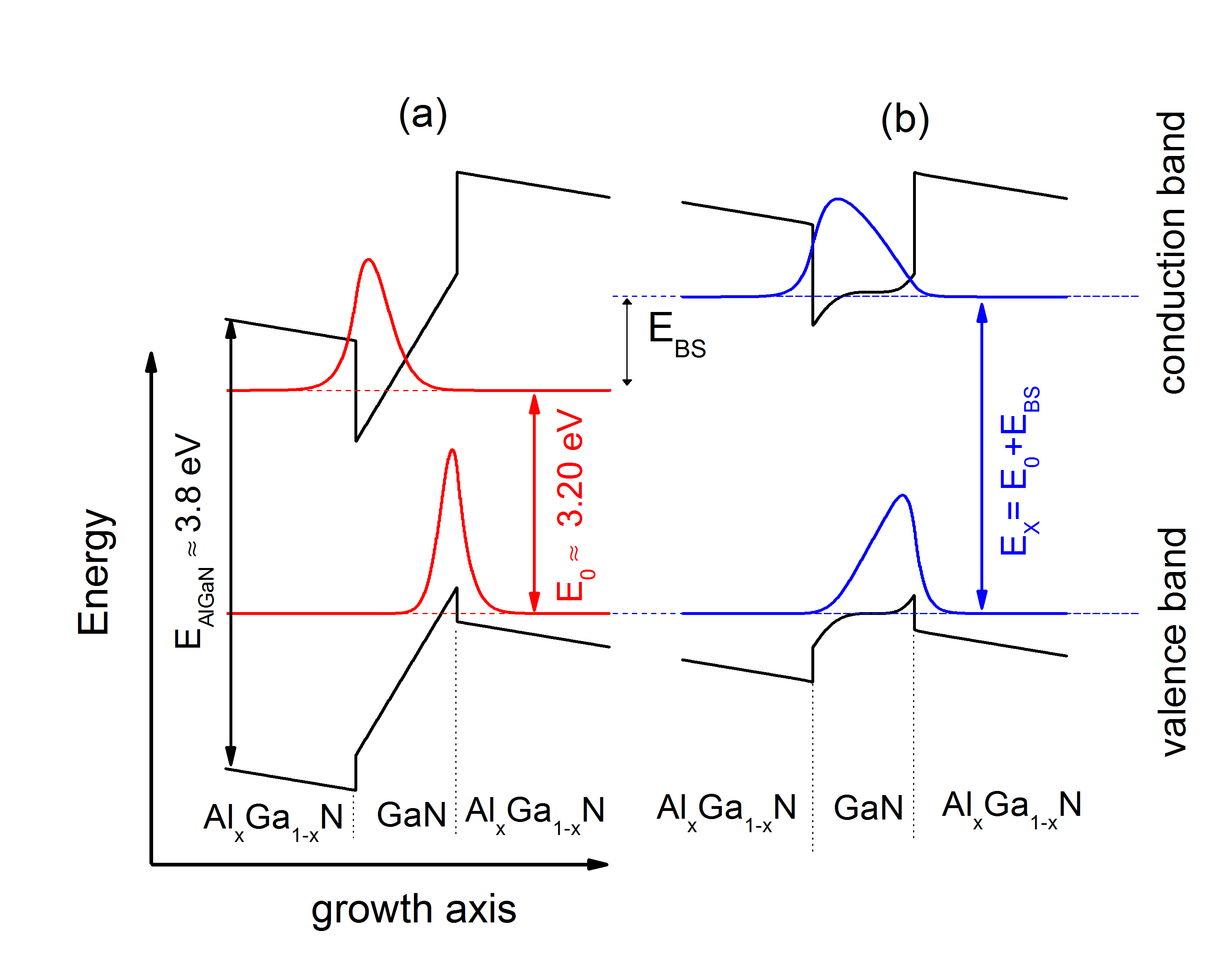}}
\caption{Calculated electronic band structures showing energy
and wave functions of electrons and holes in a polar single quantum
well (sample C) at (a) low and (b) high exciton density respectively and at $T$= 4 K.
$E_{AlGaN}$ is the energy gap in the barrier, $E_{0}$ ($E_{X}$) is the exciton recombination energy  in the zero (high) density limit, $E_{BS}$ is the population-induced blue shift $E_{BS}=E_{X}-E_{0}$.
}
\label{fig:Samplestructure}
\end{figure}

Wide band-gap semiconductors are, from this standpoint,
much better candidates: binding energies as high as $25$ meV, for bulk GaN,
or $60$ meV for bulk ZnO\cite{Janotti2009RPP} permit the
preservation of excitonic properties up to room temperature.
These
compounds, in their natural wurtzite crystal form, exhibit
another interesting feature: spontaneous and piezoelectric polarizations\cite{Bernardini1997PRB,
Seo1998PRB}.
So far, this was considered detrimental for optoelectronic
applications but it can constitute a decisive advantage for creating
of cold gases of dipolar excitons.
Indeed, GaN/(Al,Ga)N or
ZnO/(Zn,Mg)O single QWs  grown along the ($0001$) crystal axis,
naturally exhibit built-in electric fields of hundreds of kV/cm.
This results from the difference of total polarizations between the
well and barrier materials.\cite{Morhain2005PRB, Leroux1998PRB}
For
instance, in GaN/(Al,Ga)N QWs wider than $3$ nm, the
quantum-confined Stark effect is so strong that it dominates over
the quantum confinement. In this case, the
ground-state exciton energy
lies below the band gap of GaN ($3.48$ eV at $4$ K).\cite{Grandjean1999APL}

In such QWs, IXs are naturally created in the absence of an external
electric field. Their binding energy decreases significantly when
the well thickness increases but still remains largely above $10$
meV.\cite{Morhain2005PRB,Bigenwald} Consistently, radiative
lifetimes increase exponentially and values as large as tens of
microseconds have been measured for QWs widths of $6-8$ nm
only.\cite{Morhain2005PRB, Lefebvre2004PRB, Morel2003PRB} One
important consequence of the large band offsets, lifetimes and
electric field effects in polar GaN/(Al,Ga)N QWs is the possibility
to accumulate dipolar exciton densities larger than $10^{12}$
cm$^{-2}$ using reasonably small laser power densities, inducing
blue shifts of several hundreds of meV. This situation has been both
observed experimentally and thoroughly
modeled,\cite{Lefebvre2004PRB} as shown in
Fig.~\ref{fig:Samplestructure}.
By solving self-consistently the
Poisson and Schr\"odinger equations, the curvatures of both
potential and envelope wave-functions can be calculated, to obtain
the optical transition energies.

The purpose of this paper is to exploit dipole-dipole repulsion to study
the transport of IXs in GaN/(Al,Ga)N QWs grown along
the ($0001$) axis. The tightly localized, strong nonresonant excitation
in such structures produces a potential that pushes the dipolar excitons away from the excitation spot. Time and spatially-resolved photoluminescence (PL)
measurements allow us to observe the emission/propagation of those
excitons over tens of micrometers along the sample plane, at low
temperature ($T=4$ K).
When the temperature is increased, the dipolar exciton propagation exhibits a drastic
shrinking of extension, in the form of a horizon which corresponds to the
distance at which the radiative recombination rate becomes smaller than the nonradiative one.
Fittings of the experimental results in the framework of the drift-diffusion model suggests
the domination of diffusion over the drift induced by dipole-dipole repulsion.

\section{Samples}
\label{sec:samples}
\renewcommand{\arraystretch}{1.4}
\begin{table}[h]
\begin{center}
\begin{tabular}{|c|c|c|c|}
\hline
Sample name& A  & B  & C  \\ \hline
QW width (nm) & \textbf{$6.0$} & \textbf{$6.0$} & \textbf{$7.8$} \\ \hline
Al (\%) & \textbf{$20$} & \textbf{$20$} & \textbf{$15$} \\ \hline
Inner barrier & Al$_{0.2}$Ga$_{0.8}$N & AlN & - \\ \hline
F (kV/cm) & \textbf{$900$} & \textbf{$900$} & \textbf{$670$} \\ \hline
$\tau_{rad}^0$ ($\mu$s) & \textbf{$22.5$} & \textbf{$24.0$} & \textbf{$1.6$} \\ \hline
$\phi_{0}$ (eV $\times$ cm$^{-2}$) & \textbf{$6.5 \times 10^{-14}$} & \textbf{$7.49 \times 10^{-14}$} & \textbf{$5.39 \times 10^{-14}$} \\ \hline
$\gamma$ (cm$^{-2}$) & \textbf{$9.08$ $\times$ $10^{11}$} & \textbf{$11.6$ $\times$ $10^{11}$} & \textbf{$5.68$ $\times$ $10^{11}$} \\ \hline
$E_{0}$ (eV) & \textbf{$3.08$} & \textbf{$3.13$} & \textbf{$3.20$} \\ \hline
\end{tabular}
\caption{Sample parameters: GaN QW width, Al content in GaAlN barriers,
inner barrier composition, built-in electric field $F$.
The values of $\tau_{rad}^0$, $\gamma$ ($E_0$, $\phi_0$) are used to calculate density-dependent
radiative lifetimes $\tau_{rad}$ (exciton energies $E_X$),
according to Eq. \ref{eq:taurad} (Eq. \ref{eq:ex}).
}
\label{tab:samples}
\end{center}
\end{table}
The three samples under study were grown by molecular beam epitaxy
(MBE) on c-plane oriented sapphire substrates followed by a $2~\mu$m
thick GaN buffer layer. Their main characteristics are summarized in
Table \ref{tab:samples}. For sample A, the active zone consists of a
$30$ nm thick Al$_{0.20}$Ga$_{0.80}$N barrier, a $4.4$ nm ($17$
atomic monolayers) GaN QW followed by one monolayer of
Al$_{0.20}$Ga$_{0.80}$N, a $1.6$ nm ($6$ atomic monolayers) GaN QW
and a $30$ nm thick Al$_{0.20}$Ga$_{0.80}$N cap layer. This sample
can alternatively be described as a $6$ nm-wide
GaN/Al$_{0.20}$Ga$_{0.80}$N QW, in which an ultrathin barrier of
Al$_{0.20}$Ga$_{0.80}$N has been placed at the location where the
envelope functions of the electron and the hole overlap most weakly.
In such structures, the overlap is further decreased and the
radiative lifetime is enhanced, but with a limited impact on the
exciton binding energy. Sample B was fabricated following the same
concept of a $6$ nm-wide QW containing a $1$-monolayer barrier, but
the latter was made of AlN, instead of Al$_{0.20}$Ga$_{0.80}$N. For
sample C, the active zone consists of a $30$ nm thick
Al$_{0.15}$Ga$_{0.85}$N barrier, a $7.8$ nm ($30$ atomic ML) GaN QW
and $30$ nm thick Al$_{0.15}$Ga$_{0.85}$N cap layer. Sample C was
also studied in Ref.~\onlinecite{Lefebvre2004PRB}. Its
low-temperature exciton lifetime was determined as
$\tau_{rad}^0=1.65$ $\mu$s. For samples A and B, using similar
experimental conditions as in Ref.~\onlinecite{Lefebvre2004PRB} and
a laser repetition rate of $4$ kHz, we measured characteristic decay
times of $22.5$ and $24$ $\mu$s respectively. As for Sample C
(Ref.~\onlinecite{Lefebvre2004PRB}), the PL spectra of Samples A and
B peak at much lower energies than the bulk GaN exciton. They also
exhibit intense optical-phonon replica, which is another consequence
of the strong on-axis dipole.\cite{Kalliakos2002PSS, Morhain2005PRB,
Zhang2001JOP}

\subsection{Excitonic energy and radiative lifetime versus excitonic density}
\label{section:energy_vs_density}
\begin{figure}[ht!]
\center{\includegraphics[width=1.0\linewidth]{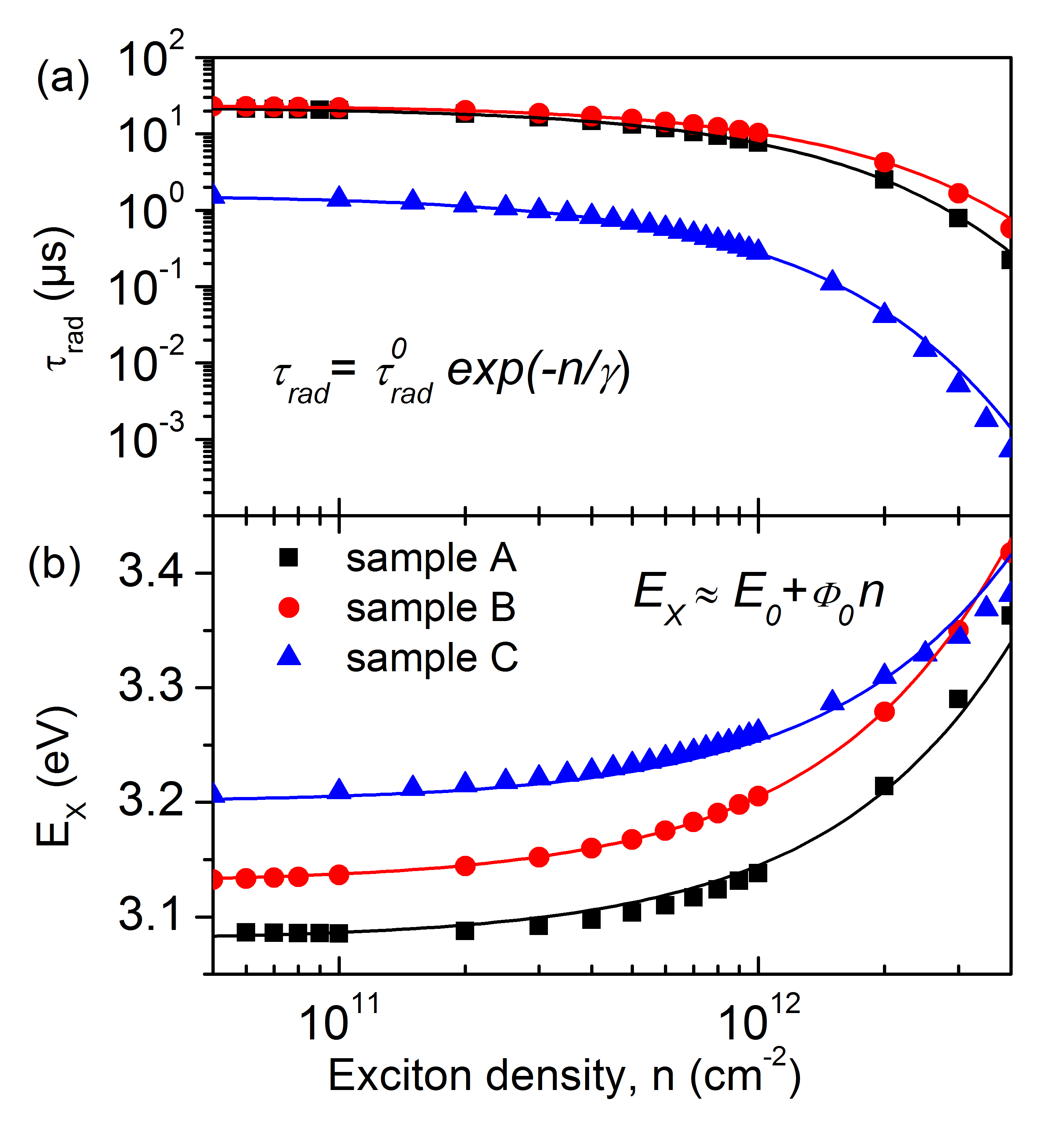}}
\caption{Calculated dependence of radiative lifetime (a) and transition energy (b) {\it versus}  exciton density,
for samples A, B and C found by solving Schr\"odinger-Poisson equations (triangles, circles and squares, respectively). Continuous lines show the approximations by exponential (lifetimes: a) and linear laws (energies: b).}
\label{fig:energy}
\end{figure}

To describe the transport of indirect excitons quantitatively, we need
to establish a correspondence between the density of dipolar
excitons and the blue shift of the optical transition. For this purpose, we
solve numerically the coupled Schr\"odinger-Poisson equations by the
self-consistent procedure described in Ref.~\onlinecite{Lefebvre2004PRB}.
Above a Mott density of $n \approx 10^{12}$ cm$^{-2}$, many body
effects and band gap renormalization are known to influence the
density induced blue shift, as explained in Ref.
~\onlinecite{Lefebvre2004PRB}, introducing a simplified,
phenomenological model for the additional terms to the field
screening effects. Recently, a more thorough discussion and modeling
of those many-body interactions has been published.
\cite{Rossbach2014PRB} In the measurements reported below, we will
be only weakly concerned by densities exceeding $10^{12}$ cm$^{-2}$.
Thus, we use the phenomenological approach of Ref.
~\onlinecite{Lefebvre2004PRB} to establish the density dependence of
the blue shift. A typical result is shown in Figure
\ref{fig:Samplestructure}, for low (a) and high (b) exciton
densities. The blue shift $E_{BS} $ from the zero-density exciton
transition $E_{0}$ can be observed at high density, as well as the
increasing overlap of the electron and hole wave functions. Note
also the strongly curved potential profile. The local electric field
is not constant along  the growth axis, making our calculation far
more accurate than others which assume a constant electric field,
rigidly altered by static charges distributed as Dirac functions at
each interfaces. The values of the built-in electric field $F$ at
zero exciton density are given in Table \ref{tab:samples} for each
sample.

Using this approach, we calculate the blue shift $E_{BS}$ of
the optical transition as a function
of the exciton density $n$ for each sample (Fig.~\ref{fig:energy}(b)).
The dependence between the two quantities is nearly
linear and the exciton energy can be approximated by:
\begin{equation}
E_{X}=E_{0}+ E_{BS} \simeq E_0+ \phi_0 n,
\label {eq:ex}
\end{equation}
where the self-induced potential $\phi_0$ is the fitting parameter.
Thus, from the experimentally determined PL blue shift
we can deduce the IX density.

The radiative lifetime $\tau_{rad}$ is inversely
proportional to the squared overlap integral of the electron
and hole envelope functions, and can be calculated as a function of the exciton density (Fig.~\ref{fig:energy}(a)).
The resulting density-dependent lifetime for all the samples is well described by an exponential
\begin{equation}
\tau_{rad}\simeq\tau_{rad}^0 \exp(-n/\gamma),
\label {eq:taurad}
\end{equation}
where $\gamma$ is the fitting parameter.
$\tau_{rad}^0$ is measured by  macro-PL measurements of the ultimate decay time constants,
measured at low temperature and at long delays. The resulting parameters are reported  in Table \ref{tab:samples} for the three samples.

\begin{figure*}[t]
\center{\includegraphics[width=1.8\columnwidth]{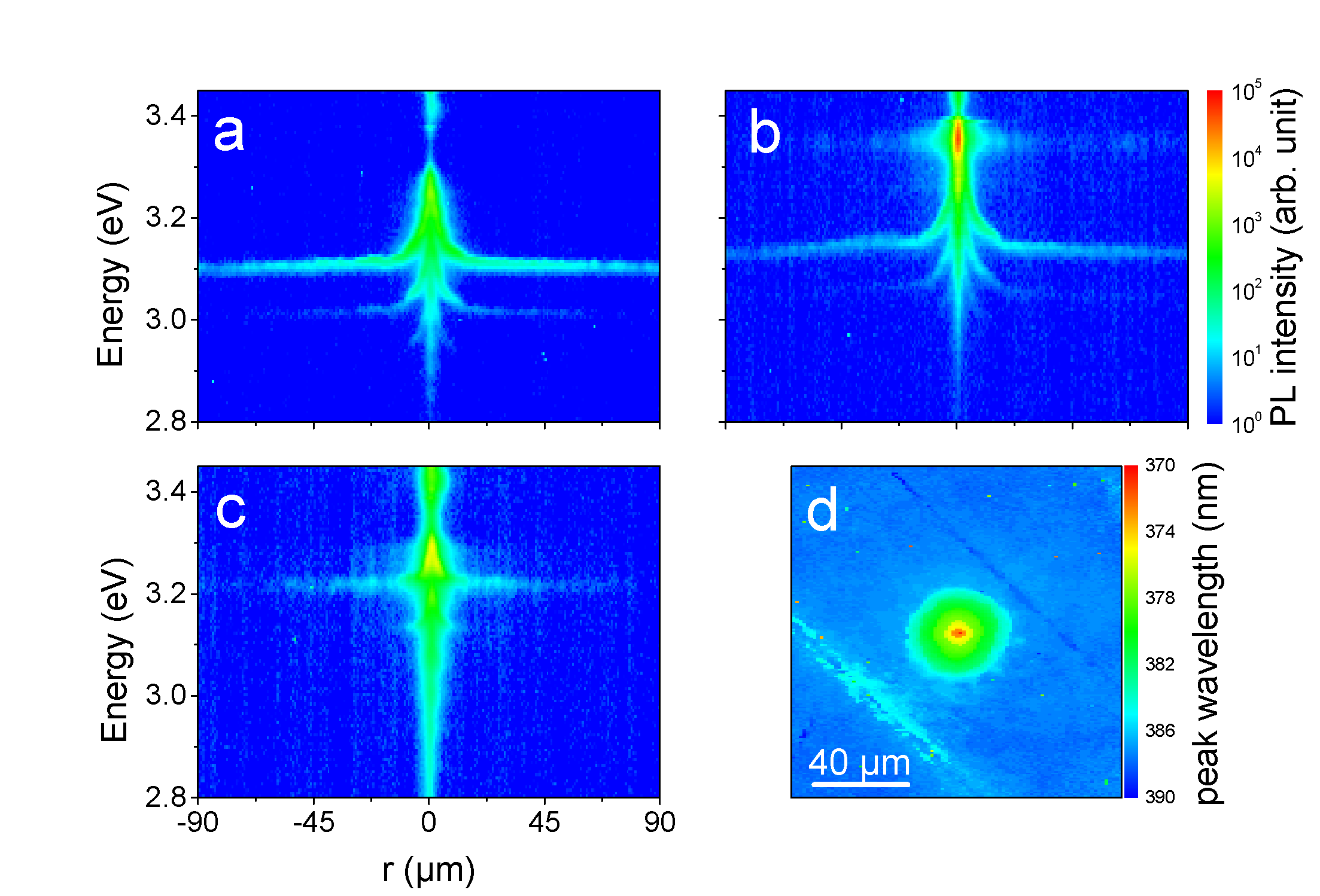}}
\caption{Color map of the PL intensity taken at $T=4$~K on (a) sample A, (b) sample B, and (c) Sample C. The abscissa $r$ corresponds to the distance from the excitation spot. The ordinate corresponds to the energy spectrum. The color map (d) shows the PL peak wavelength as a function of the position in the ($x,y$) plane of Sample B. The excitation spot is located at the center of the map.}
\label{fig:allsamples}
\end{figure*}

\begin{figure*}[ht!]
\center{\includegraphics[width=1.6\columnwidth]{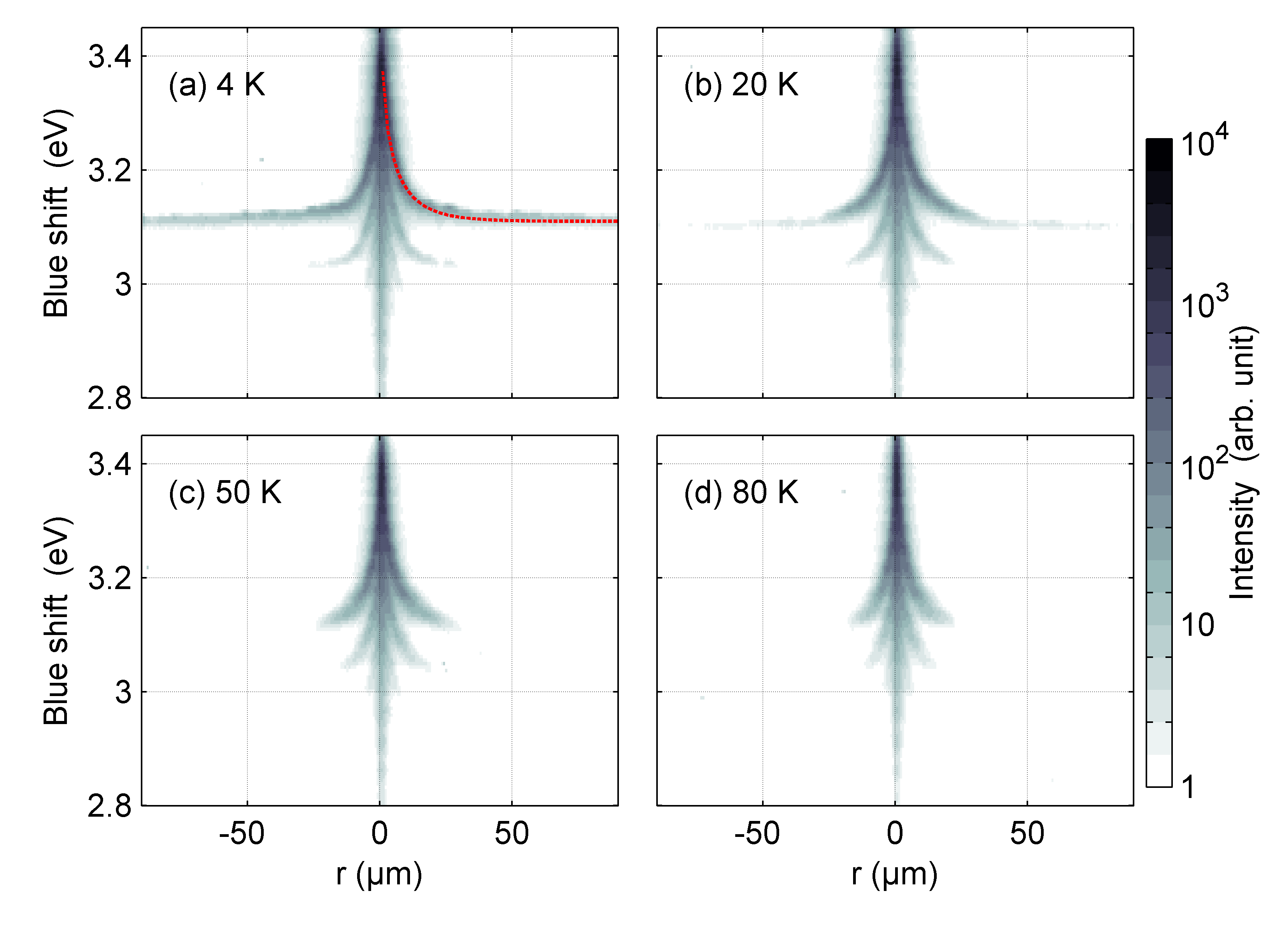}}
\caption{Grayscale map of the PL intensity taken on sample A for different temperatures: (a) $4$~K, (b) $20$~K, (c) $50$~K and (d) $80$~K. The solution of the diffusion equation with constant diffusion coefficient $D$ and recombination rate $R$ such that
$\sqrt{D/R} = 12$~$\mu$m is displayed in (a) as a red dotted line.}
\label{fig:sampleA}
\end{figure*}
\section{Experimental setup}
For micro-PL experiments, the samples were placed in a
cryostat and excited by a $266$ nm continuous-wave
laser (Crylas/FQCW $266-50$) through a microscope objective.
This objective collects the PL signal in such a way that the image
of the sample surface  is reproduced at the entrance of the spectrometer, where
it is filtered by a vertical entrance slit.
The spectrometer is
equipped with a grating with $150$ gr/mm blazed at $390$ nm.
The detector is a cooled CCD camera with $1024 \times 256$ pixels. 
The pixel size is $26$ $\mu$m, which corresponds to $760$ nm on the sample surface.
This value matches the smallest laser spot diameter that our objective can provide ($800$~nm),
thus allowing for an overall spatial resolution of $0.8$~$\mu$m.
This setup provides a direct spatial imaging of the luminescence with a spectral resolution.
%
Two-dimensional images, with
spectral resolution, were obtained by scanning horizontally the lens
which ensures the focusing onto the same entrance slit. The
excitation power density was estimated as P$_{0}= 2.3$~MW/cm$^{2}$.

PL experiments combining time and spatial resolution are based on the same setup,
except that the detector is a Hamamatsu streak-camera (model C$10910$ equipped with an S$20$ photocathode and UV coupling optics), enhanced for ultraviolet detection. This camera can reach an ultimate time-resolution of $1$ ps but, for the time-windows explored in this study, this resolution was rather of $15$~ps. Pulsed laser excitation at $260$ nm, through the same microscope objective, is provided by the frequency-tripled tunable
radiation from a Titanium-Sapphire cavity, with typical pulse duration of $150$ fs
and tunable repetition rate. For the present ultra-slow decays, we set this rate to $4$ kHz with an average power density 0.1 MW/cm$^2$.
\section{Experimental results}
\label{sec:ER}

\subsection{Low-temperature PL}

Figure~\ref{fig:allsamples}(a-c) shows typical images of spatially
and spectrally resolved PL  obtained at T = $4$ K for Samples A, B
and C. The PL intensity is measured at different distances $r$ from
the excitation spot (horizontal axis) and  at different photon
energies (vertical axis), in the presence of the strongly focused
excitation at $r=0$ at power density $P_0$. The dominant PL line, at
higher energy, corresponds to the zero-phonon excitonic
recombination. Optical phonon (LO) replicas of this PL line appear
at lower energies, separated by multiples of $91$ meV. As far as the
spatial dependence is concerned, these replica follow the energy
variation of the main zero-phonon line. In the following we will
focus on the zero-phonon emission line. Close to the excitation
spot, the PL energy is blue shifted up to values close to the GaN
excitonic gap $E_{GaN}=3.48$ eV. The maximal blue shift, observed at
$r \sim 0$, is larger than $0.2$ eV for Samples A and B, see Figure
\ref{fig:allsamples}(a-b), and close to $0.15$ eV for Sample C, see
Figure~\ref{fig:allsamples}(c).
However, the determination of the exact blue shift at $r<3$~$\mu$m
is less accurate than for the rest of the sample surface, because
the observed PL spectra are quite broad, owing to multi-excitonic
effects, and other many-body interactions. \cite{Rossbach2014PRB}
At  $r < 20$ $\mu$m, the $r$-dependence of the PL maximum has the
shape of an arrow, characterized by a decrease of the blue shift
accompanied  with a decrease of intensity (note the logarithmic
color-encoding of the intensity). This emission profile will be
referred to as the arrow-shaped pattern.

At longer distances ($r > 20$ $\mu$m), the PL energy  is nearly constant,
but the signal persists at surprisingly long distances.
For Samples A and B, we observe excitonic emission up to $100$ $\mu$m away from the generation spot.
For Sample C, with the smallest initial energy shift, the excitonic emission is still observable up to several tens
of $\mu$m away from the excitation spot.

Figure~\ref{fig:allsamples}(d) displays a color map
of the QW peak emission wavelength as a function of the real-space coordinates
($x,y$) on the surface of Sample B. The above-mentioned central arrow-shaped
region appears to be of circular symmetry, with a radius
of $20$ $\mu$m. Then the PL peak wavelength remains nearly constant over
distances as large as $100$ $\mu$m. Note that this wavelength is slightly
altered by the presence of linear microscopic defects (cracks),
most certainly owing to local variation of
barrier composition and/or strain field, near
such defects. Careful examination of Figure~\ref{fig:allsamples}(a-c) also
emphasizes small but reproducible fluctuations of both
intensity and PL energy shift, when moving away from the excitation spot.
We attribute these features to local fluctuations of barrier composition and/or
strain, which modify the local accumulation of excitons and therefore their
energy.

\subsection{Temperature dependence of continuous-wave PL}
Figure~\ref{fig:sampleA} shows the  temperature dependence of the spatially-resolved PL spectra
for Sample A, obtained at power density P$_{0}$.
%
One can observe the shrinkage of
the spatial extension of the PL, when the temperature is increased.
The PL signal falls below the detection threshold of the setup above a critical distance.
This critical point can be referred to as an emission horizon, and it
gets closer to the excitation spot as the temperature is increased.
\begin{figure}[ht!]
\center{\includegraphics[width=1\linewidth]{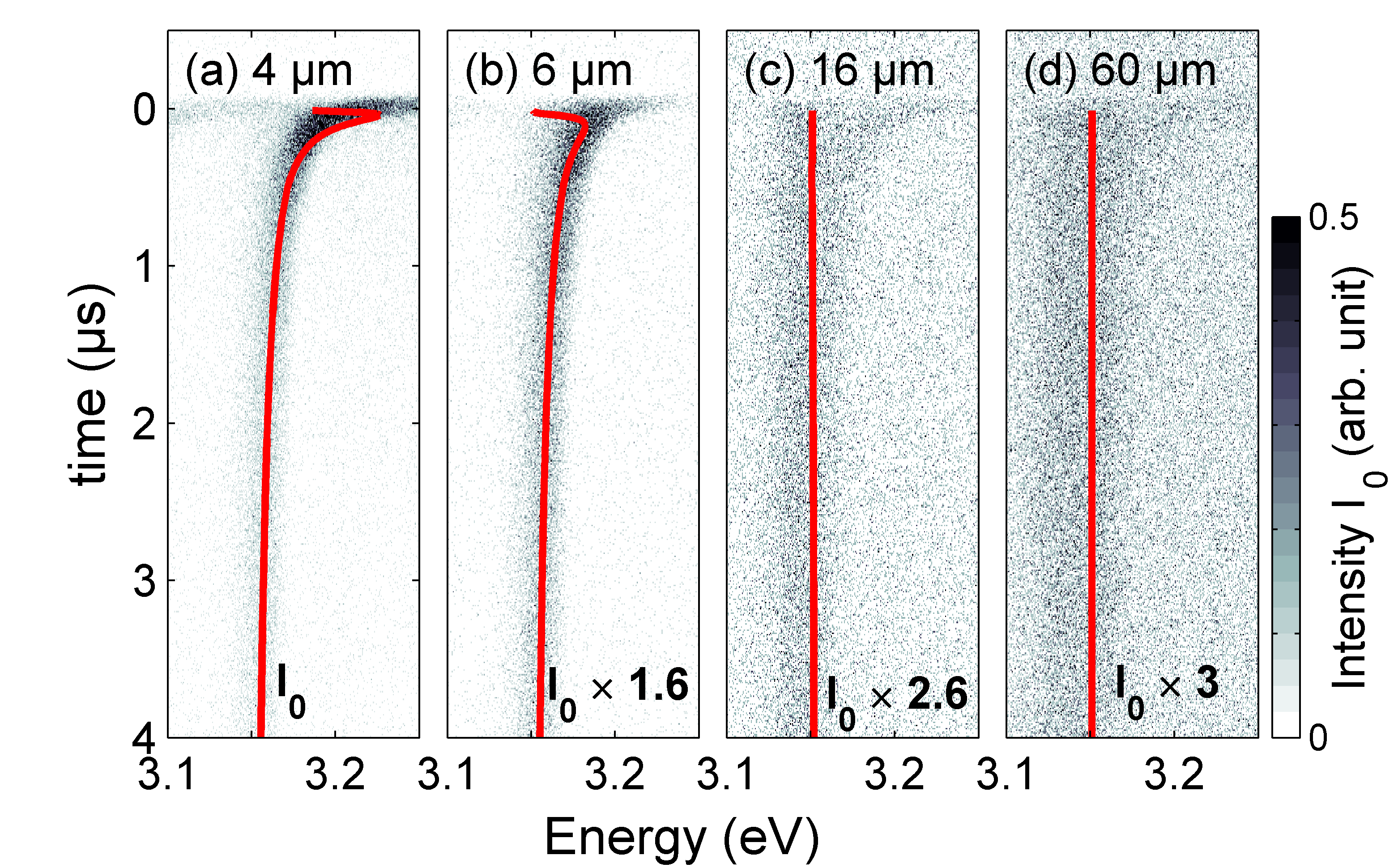}}
\caption{Time evolution of the PL intensity $I_0$ for sample A, taken at $T=4$~K and at
(a)$4$ $\mu$m,
(b)$6$ $\protect\mu$m,
(c)$16$ $\protect\mu$m and (d): $60$ $\protect\mu$m away from the excitation spot.
The grayscale map indicates the intensity of emission at a given energy (horizontal scale) and at a given time (vertical scale). Red curves show the calculated change of transition energy, with time, for each distance.}
\label{fig:timeresolved}
\end{figure}

\subsection{Time-resolved experiments}

To get a deeper insight into the dynamics of exciton propagation, we performed
low-temperature time-resolved PL experiments. By conjugating the temporal slit
of the streak-camera to a given point on the surface sample, we are able to
record the PL intensity collected at a given energy and at a given delay.
Figure \ref{fig:timeresolved} shows examples, in the case of Sample A,
of such time-dependent PL spectra, gathered at $4$ different distances
from the excitation spot. The time-domain has been restricted, here, to the
first $4$ microseconds of decay, but we have also measured such decays over
a full time range of $50$ $\mu$s, and we have examined, too, the first
$500$ ns after the excitation time.

In the vicinity of the excitation spot (Figure \ref{fig:timeresolved}(a)),
the observed dynamics is the same as in  spatially integrated
experiments:\cite{Lefebvre2004PRB} less than 50~ns after the excitation pulse,
there is a sudden blue shift of the PL emission which results from the fast
creation of dipolar excitons. Then, the exciton population decays and the PL
peak energy decreases, until it reaches a steady value.
Only in the latter regime, the intensity time-decay becomes exponential
and provides us with the decay times that we gathered in Table \ref{tab:samples}.
As we increase the distance between the observation and the excitation spots
(Figure \ref{fig:timeresolved}(b)),
we observe the same kind of behavior, but with a smaller initial blue shift:
lower exciton densities are reached, consistent with the continuous-wave PL results.
It can still be argued that the time needed to propagate over $6$ $\mu$m is shorter
than the time resolution available in these experiments.
However, the most puzzling results are those obtained above $15$
$\mu$m (Figure~\ref{fig:timeresolved}(c-d)). Indeed, even at these
large distances, the PL signal builds up within a few nanoseconds
after laser excitation. This observation proves that, for distances
larger than $15$-$20$ $\mu$m, the observed exciton recombination can
in no way arise from the propagation of those excitons, at least in
the sense of a classical drift/diffusion process. As discussed in
the next section, no reasonable diffusion coefficient can explain
such a fast arrival of excitons far away from the excitation spot.
However, the quenching of the PL tail observed at high temperatures
indicates the excitonic origin of this signal. We suggest that this
observation results from the direct optical creation of secondary
excitons, probably by the intense luminescence emitted under the
laser spot and guided along the sample plane.

%

\section{Discussion and modeling}

\subsection{Origin of the arrow-shaped micro-PL pattern.}

The arrow-shaped pattern apparent in Fig. \ref{fig:allsamples}(a-c)
is a signature of the decreasing density of excitons,
when the distance from the excitation spot increases.
The dipole-dipole repulsion between excitons pushes them away from the excitation spot.
Exciton propagation, accompanied by recombination, results in a decreasing exciton density.
%
Generation, diffusion, drift, localization, radiative and
nonradiative recombination are the competing mechanisms that we need
to unravel, in order to model quantitatively our observations.
We
will assume that the blue shift directly reflects the local exciton
density, whereas the PL intensity also results from the competition
between radiative and non-radiative recombination processes.
 The
spatial dependence of the blue shift, observed less than 20~$\mu$m
away from the excitation spot, appears to be  temperature
independent below $T \le 50$~K, within our experimental resolution.
Since drift mechanisms are typically temperature dependent because
of the Einstein relation, we can first infer that either the
thermalization is not complete below $T$=50~K for short propagation
distances, or that the exciton drift plays only a minor role here,
due to the complex multiscale disorder potential.

The simplest way to reproduce the arrow-shaped diffusion pattern is
to consider the diffusion of excitons generated by a delta-peaked
excitation pulse at $r=0$. Let us assume a constant diffusion
coefficient $D$ and a constant recombination rate $R$. Far from the
excitation spot, the steady state diffusion equation is reduced to
$D \Delta n =R n $. This equation has an analytical solution:
$\phi_0 n$ = $A \times {\cal{K}}_0 (r \sqrt{R/D})$, where $A$ is a
constant and ${\cal{K}}_0$ is the modified Bessel function of the
second kind. This solution with $A=0.1$ eV and $\sqrt{D/R} \sim
12~\mu$m is shown by the red curve in Fig.~\ref{fig:sampleA}(a) that
we superposed onto the PL map. Even though we know that this model
is based on the un-satisfied assumption of a constant exciton
lifetime and a constant diffusion coefficient, it
reproduces reasonably the overall shape of the PL spatial dependence. 
\subsection{Origin of the PL tail}
A model based on purely diffusive propagation of IXs
can, in principle, reproduce both the PL arrow-shape and the absence of
temperature dependence, for $r < 20~\mu$m. On the other hand,
the tail of the PL line, for $r > 20~\mu$m cannot be reproduced
by diffusion of excitons, because it would imply an anomalously high diffusion constant.
Indeed, time-resolved measurements of PL at 60~$\mu$m from the
excitation spot indicate that excitonic emission builds up in less
than 100~ns after the excitation pulse (Fig.~\ref{fig:timeresolved}(d)).
This cannot either be reproduced by the exciton drift, because of the weak exciton density gradient observed at $r > 20~\mu$m. We therefore suggest that a significant part of the higher-energy photons emitted around $r = 0$ by the QW itself is guided along the sample plane and is capable of creating, at larger distances, secondary excitons.
Let us now address the temperature dependence of this emission and the reduction of the horizon.
First of all, a common property of III-nitride-based heterostructures is a high density of nonradiative recombination centers (which are mainly threading dislocations), of the order of $10^{10}$ cm$^{-2}$. This is the case in the present heteroepitaxial QWs.
Therefore, as the temperature delocalizes IXs,  it also enhances the rate of their capture by those nonradiative traps. Given the very large values of the radiative lifetimes in our samples, at least in the unscreened regime at $r > 20~\mu$m, the competition between slow radiative and faster nonradiative recombinations explains why PL signals are quenched, when the temperature is increased, for distances exceeding a certain horizon. This horizon is simply determined by the exciton nonradiative recombination rate, which increases with the temperature.
\begin{figure*}[ht!]
\center{\includegraphics[width=1.6\columnwidth]{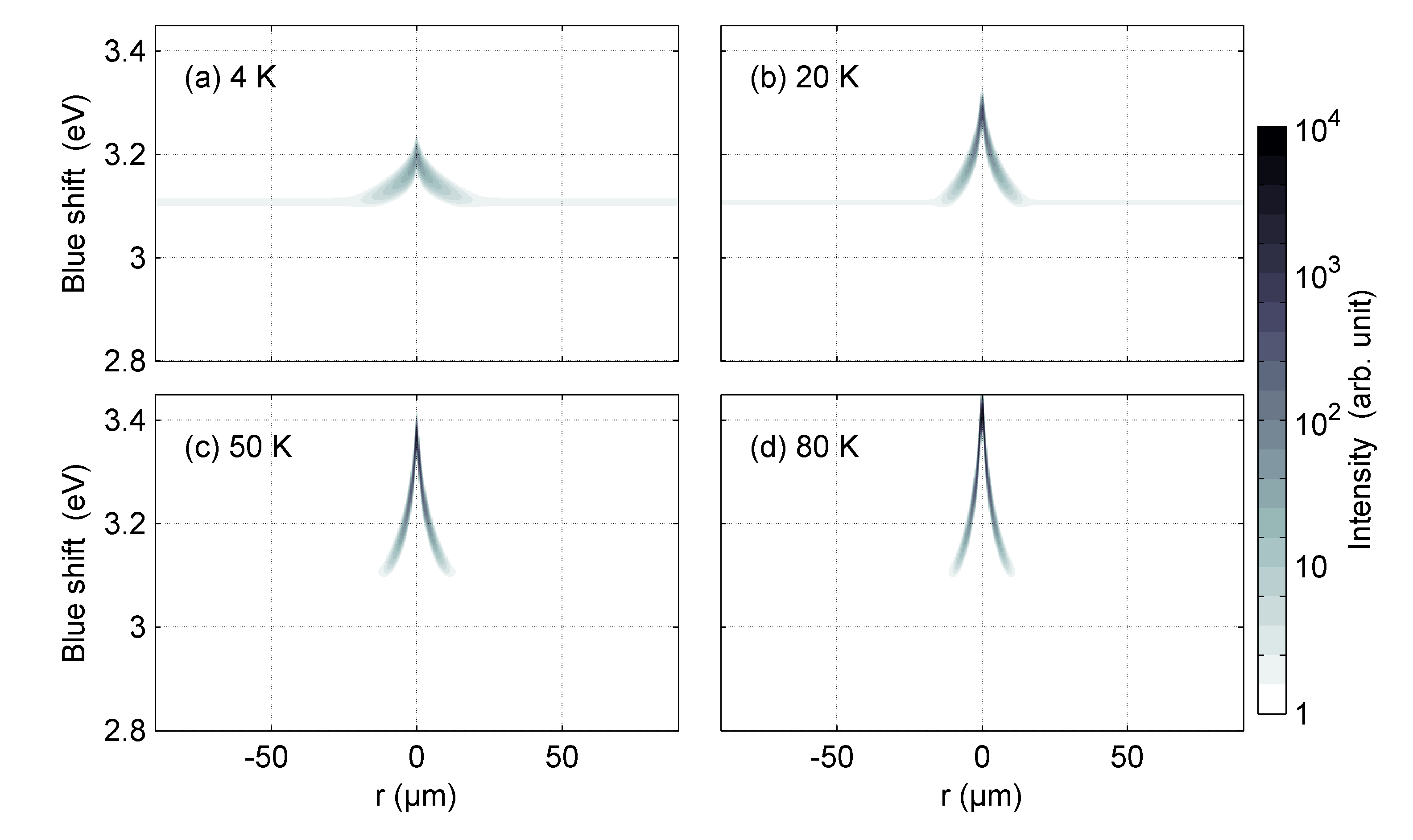}}
\caption{Grayscale map of the PL intensity calculated using Eq. \ref{eq:drift-diffu} at different temperatures: 4, 20, 50 and 80 K as a function of the distance $r$ from the excitation spot, and emission energy. The drift current given by Eq. \ref{eq:Jdrift} is taken into account. A phenomenological Gaussian broadening of the blue shift with a standard deviation of 70 meV has been introduced.}
\label{fig:PL_with_drift}
\end{figure*}

\subsection{The transport equation}
In this section, we describe a formal model for the exciton propagation.
In first approximation we neglect the thermalization effects and
consider that excitons are close to the lattice temperature as soon as they leave the excitation spot.
 The equation for the in-plane transport of indirect excitons is:\cite{Ivanov2006}
\begin{equation}
\frac{\partial n}{\partial t}=
- \nabla \cdot \mathbf{J} + G - R n,
\label{eq:drift-diffu}
\end{equation}
where $n$ is the exciton density, $G$ and $R$ are the generation and recombination rates, respectively, $\mathbf{J}$ is the IX current density, which can be split into drift and diffusion components:
$\mathbf{J}= \mathbf{J}_\mathrm{drift} + \mathbf{J}_\mathrm{diff}$.

The diffusion current $\mathbf{J}_\mathrm{diff}$ is given by:
\begin{equation}
\mathbf{J}_\mathrm{diff}=
-D \nabla n
\label {eq:diffcurr}
\end{equation}
where $D$ is the exciton diffusion coefficient.
The random disorder potential, caused by alloy disorder in the barrier
and by QW width fluctuations, hinders the in-plane exciton
propagation along the GaN/AlGaN QW plane.\cite{Gallart2001PSSA}
For high exciton densities, however, the potential fluctuations
are screened and excitons, on average, become more mobile. This effect can
be taken into account in the framework of a thermionic model,\cite{Ivanov:2002}
which gives the following density dependence:
\begin{equation}
D= D_0 \exp \left(
\frac{-U_0}{\phi_0 n + k_B T}
 \right) ,
 \label {eq:D}
\end{equation}
where $D_0$ is the diffusion constant in the absence of disorder and
$U_0$ is the amplitude of the disorder potential at zero density.
This formulation implicitly accounts for the modification of the
fluctuation potential by screening of the internal electric field.
\cite{Ivanov:2002}
 We estimate $U_0 \simeq 30$~meV, which corresponds to the energy shift induced by a
fluctuation of one monolayer in the QW width.
$D_{0}$ represents the free exciton bulk diffusion coefficient,
which is reached when the disorder potential $U_{0}$ is negligible compared
to the blue shift $\phi_0 n$ and/or the thermal energy $k_B T$.

\begin{figure*}[ht!]
\center{\includegraphics[width=1.6\columnwidth]{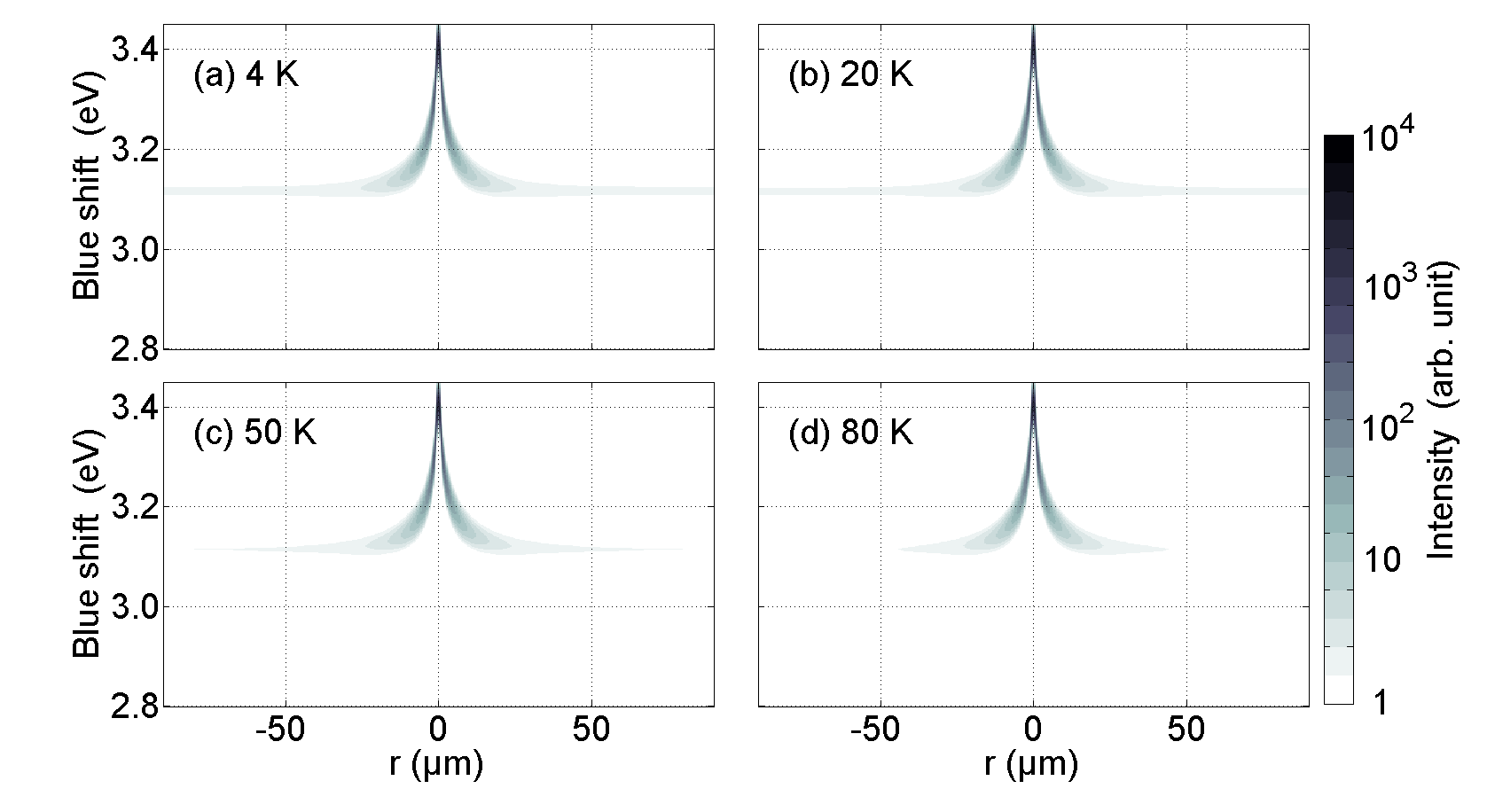}}
\caption{Grayscale map of the  PL intensity, calculated as in Fig.~\ref{fig:PL_with_drift}
except that only the diffusion current $J_\mathbf{diff}$ (Eq.~\ref{eq:diffcurr}) is taken into account
whereas the drift current $J_\mathbf{drift}$ (Eq.~\ref{eq:Jdrift}) is neglected.
}
\label{fig:modellingSR}
\end{figure*}

The drift term $\mathbf{J}_\mathrm{drift}$ in Eq.~\ref{eq:drift-diffu} can be tentatively
rewritten as:
\begin{equation}
\mathbf{J}_\mathrm{drift}=
-\mu n \nabla (\phi_0 n)
\label{eq:Jdrift}
\end{equation}
where the exciton mobility $\mu$ is connected to the diffusion coefficient $D$
via the  Einstein relation:
\begin{equation}
\mu=\frac{D}{k_{B}T}.
\label{eq:einstein}
\end{equation}
In Eq.~\ref{eq:Jdrift}, the drift is governed by the self-induced potential $\phi_0 n$ only.
We expect this relation to be valid only when $U_0 \ll \phi_0 n$, {\it i.e.} when the disorder potential is efficiently screened.

The generation term $G$ which appears in Eq.~\ref{eq:drift-diffu} will be divided into two parts:
$G= G_0+ G_{bg}$.
The first term $G_0$ corresponds to the exciton generation under the laser spot:
$$
G_0= \left(2 N_p / R_0^2 \right)  \exp (-r^2/R_0^2),
$$
where $r$ is the distance from the center of the excitation spot, $R_0$= 500 nm
is the spot radius, $N_p$ is the exciton generation rate.
The second term $G_{bg}$ is semi-phenomenological. We introduce it because any attempt to model the ultra-fast appearance of exciton recombination at $r > 20$ $\mu$m within a drift-diffusion model using reasonable diffusion coefficients is doomed to fail. Our current understanding of this phenomenon is that a significant part of the higher-energy photons emitted around $r = 0$ are most certainly guided along the sample plane, between the QW and the surface. Despite small absorption coefficients, some of those photons can be re-absorbed in the regions of $r> 20$ $\mu$m and recycled, by thermalization, into low-energy excitons, with densities comparable to, or lower than $10^{11}$ cm$^{-2}$. This explains the constancy of emission energy versus $r$, whereas the nearly constant intensity, for $20 <$ $r$ $< 100$ $\mu$m, is consistent with the weakness of the absorption coefficient, {\it i.e.} with the vanishing attenuation of the excitation.
We therefore introduce a weak background source term $G_{bg}$ to mimic this photon-recycling process. For simplicity, we take it independent of $r$.

The recombination rate $R$ can be expressed as a function of the radiative and nonradiative recombination times $\tau_{rad}$ and $\tau_{nrad}$:
\begin{equation}
R= 1/\tau_{rad} + 1/\tau_{nrad}.
\end{equation}
The density dependent radiative recombination time $\tau_{rad}$ = $\tau_{rad}^0 \exp ( - n/\gamma)$
is determined according to the procedure described in the Section \ref{section:energy_vs_density}, and does not depend on the temperature.
By contrast, nonradiative time is temperature dependent. Indeed, when either exciton concentration or temperature increases, the excitons delocalization leads to more efficient capture of excitons on the nonradiative defects.
These effect gives rise to the quenching of the PL at high temperature and determine the exciton nonradiative horizon. 

According to Refs.~\onlinecite{Kagan:1963, Avrutin:1988}, the density of a exciton flux $j$ flowing through a nonradiative recombination center of size $a \ll l$, $l$ the mean free path,  is independent of the shape of the center and equal to
$j= n  v /4$,
where $v$ is the average absolute velocity of the excitons for which a uniform distribution
of the velocity direction is assumed.
Therefore, assuming that all excitons which reach the recombination center are captured,
the rate of exciton capture is proportional to
\begin{equation}
1/\tau_{nrad}= \frac{n v}{4} 2 \pi a / l_{nrad}^2
\end{equation}
where $l_{nrad}$ is the average distance between the nonradiative recombination centers.
Using the relation $D= v^2 \tau /2$, where $\tau$ is the momentum relaxation time, we finally obtain the relation:
\begin{equation}
\tau_{nrad} = \tau_{nrad}^0
\exp \left(
\frac{-U_0/2}{\phi_0 n + k_B T}
 \right) ,
 \label{eq:taunrad}
\end{equation}
where $\tau_{nrad}^0$, which we will use as an adjustable parameter,
is the nonradiative recombination time in the absence of structural disorder $U_0=0$.

As the in-plane exciton propagation  has a cylindrical symmetry centered at the excitation spot, Eq.~\ref{eq:drift-diffu} reduces to a one-dimensional partial differential equation. For the numerical resolution, we implemented a one-dimensional finite element scheme, where the non-linear Eq.~\ref{eq:drift-diffu} is replaced by a linear one by using the Euler method.\cite{FreeFem}
For the simulation of the time-resolved experiments, the temporal evolution of the exciton density is calculated within a time-implicit scheme, assuming generation rates $G_0$ and $G_{bg}$ to be non-zero only at the initial time $t=0$.
For the experiments under constant illumination,
we take $\partial n/\partial t=0$ and $G$ is time-independent. In both cases we use Neumann
boundary conditions at $r =0$ and $r = 200~\mu$m.
Finally,  the PL intensity is obtained from the exciton density as  $n / \tau_{rad}$.
\begin{table}[h]
\begin{center}
\begin{tabular}{|l|l|l|l|l|}
\hline
$~D_0~$ & ~$\tau_{nrad}^0$~ & $~U_0~$ & $N_p$ & $G_{bg}$                           \\ \hline
1       &       1         &  30   & (a) 1.2 $\times 10^4 \delta(t=0)$ & $N_p \times 10^8$ \\
        &                  &          & (b) 1.2 $\times 10^{12}$              &                                \\ \hline
\end{tabular}
\caption{Main parameters of the fit for sample A. The diffusion constant is given in cm$^2$/s, the recombination time in $\mu$s, the disorder potential in meV.
$N_p$ is  the number of exciton generated per second for (a) time-resolved experiment and (b) experiments under continuous illumination.
The background generation $G_{bg}$ is in  m$^{-2}$s$^{-1}$ and is taken proportional to $N_p$. $\delta$ is the Dirac delta function.
}
\label{tab:fitparam}
\end{center}
\end{table}

\subsection{Numerical results and discussion}

The numerical implementation of the model points out that
drift current is not the dominant transport mechanism for
excitons. In all our attempts to introduce a drift current,
increasing temperature resulted in a progressive shrinking
of the spatial extension of the calculated arrow-shaped PL,
and an increasing blue shift. Typical results of the simulation
for Sample A are reported in Fig.~\ref{fig:PL_with_drift}.
We have introduced the Gaussian broadening of the PL energy with
the standard deviation $\sigma=70$ meV, to account for the
broadened PL lines.
One can see that the arrow-shaped feature is significantly
temperature-dependent, in contrast with the experimental
result  shown in Fig.~\ref{fig:sampleA}.
Similar results have been obtained for the two other samples.

The discrepancy with the experimental data suggests that
Eq.~\ref{eq:Jdrift} strongly overestimates the drift current,
probably because the gradient of exciton density induces too
small energy variations compared to the disorder amplitude,
which can present a complex multi-scale profile.
By contrast, when only the diffusion term is included, the
simulation gives results strikingly similar to the experiments. The
solutions of Eq.~\ref{eq:drift-diffu}, assuming
$\mathbf{J}_\mathrm{drift}=0$, are presented in Figure
\ref{fig:modellingSR}. By using the set of fixed parameters gathered
in Table \ref{tab:samples} and fitting parameters given in Table
\ref{tab:fitparam} we reproduce the emission patterns observed
experimentally.
The onset of the nonradiative horizon, as the
temperature is increased, is well accounted for by the thermionic
description of the delocalization-capture of excitons by defects.

This purely diffusive model also allows us to describe the time resolved experiments,
using the same set of parameters.
 The results of the calculation
for Sample A is shown by red curves in Fig.~\ref{fig:timeresolved}.
One can see that the onset of the PL blue shift associated which exciton
diffusion from the excitation spot towards the detection point is probably
too fast to be resolved in the present experiments (Fig.~\ref{fig:timeresolved}(a,b)).
On the same time, the immediate (in less than $100$~ns) onset of the
excitonic emission at large distances from the excitation spot has
the same origin as the PL tail in the spatially resolved images. It
is  due to secondary absorption of the PL guided in the sample
plane, described by the semi-phenomenological generation term
$G_{bg}$.

The arrow-shaped patterns of PL measured and calculated here are
similar to the patterns that have been reported in GaAs CQWs under
external gate voltage \cite{Butov2002,Voros2005PRL,Ivanov2006}.
However, the typical energy and distance scale are quite different
in our structures. In particular, the energy scale of the
arrow-shaped feature reaches $200$ meV, exceeding by more than an
order of magnitude the GaAs values.
Modeling of the PL patterns within the exciton drift-diffusion model
taking into account the particularities of the nitrides, suggests
that even in the highest density and lowest temperature regime
exciton transport is dominated by density-activated diffusion,
rather than by the drift current.
This effect seems to be proper to GaN QWs, but further studies are
necessary for its better understanding.
In particular the design of heterostructures without guided PL and
the resulting  emission would be useful.
%

\begin{figure}[ht!]
\center{\includegraphics[width=1.0\linewidth]{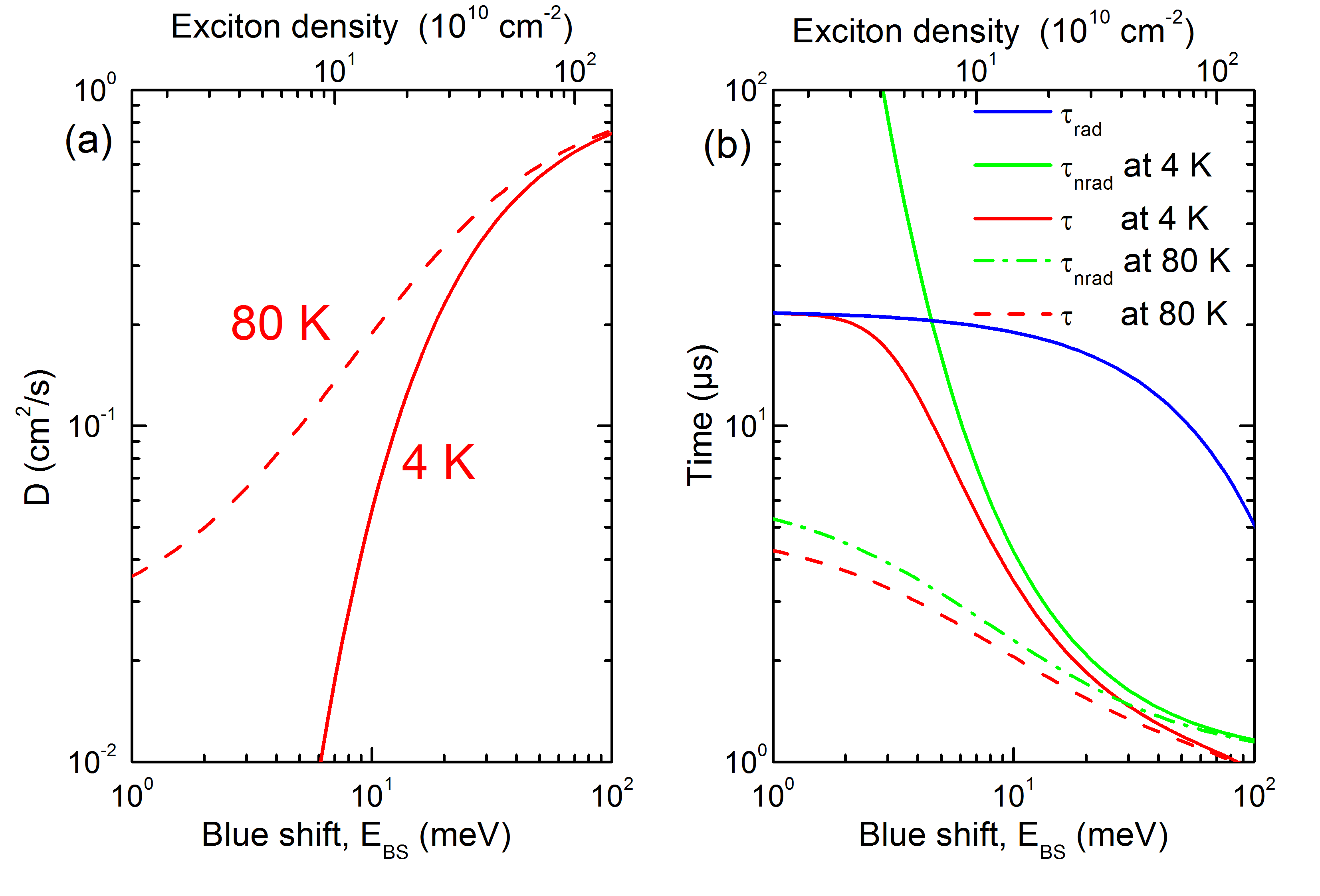}}
\caption{Illustration of the modeling assumptions for Sample A. Diffusion coefficient (a) and exciton lifetime $\tau$ (b) as a function of the blue shift at T$=4$~K (red solid line) and T$=80$~K (red dashed line). Radiative  and nonradiative components $\tau_{rad}$ and $\tau_{nrad}$ are shown by blue and green lines, respectively.}
\label{fig:TandD}
\end{figure}

It is instructive to analyze the density dependence of the diffusion coefficient (Eq. \ref{eq:D}) and recombination times (Eq. \ref{eq:taurad}, Eq. \ref{eq:taunrad}), assumed in this model. These quantities are shown in Fig.~\ref{fig:TandD} for Sample A.  The density scale (upper $x$-axis) is reported together with the blue shift scale (lower $x$-axis).
One can see that at low temperatures ($T=4$ K) the diffusion coefficient is far from being constant in the range of the experimentally observed energy shifts.
This results precisely from the dipole-dipole repulsion energy, $E_{BS}=\phi_0 n$. The diffusion becomes exponentially slow at low density and only when the disorder is screened by the dipole-dipole repulsion the diffusion coefficient  $D$ reaches asymptotically the maximum value of $D=D_0=1$ cm$^2$/s. This behavior is smeared out at higher temperatures, because the thermal energy contributes to the delocalization.
For high exciton densities, comparable to, or larger than $10^{12}$ cm$^{-2}$, the random potential fluctuations $U_0$ are overcome even at $T=80$ K, because $E_{BS}  \gg U_0, k_B T$.
This corresponds to the situation near $r= 0$ in our PL experiments, and explains why the change of the temperature has very little influence on the observations in this region. As the exciton density decreases from $10^{12}$ to $10^{11}$ cm$^{-2}$, the diffusion coefficient is reduced by two orders of magnitude, for $T= 4$~K, owing to the relocalization of carriers. This is what occurs, as the distance $r$ is increased : the situation goes continuously from the domination of mobile, delocalized excitons to frozen excitons, localized at random fluctuations. If the temperature is increased, as shown in Fig. \ref{fig:TandD}, thermal excitation takes over ($k_B T >  U_0,  E_{BS}$) and therefore the diffusion persists even at lower exciton densities.

Similarly, Fig.~\ref{fig:TandD}(b) shows the calculated
exciton lifetime $\tau$ and its radiative and nonradiative  components.
 Despite a much smaller radiative lifetime at high densities, the radiative efficiency, there, is not maximum: the delocalization of carriers favors their nonradiative capture. On the other hand, at $T= 4$~K, reducing the density dramatically increases the radiative lifetime but increases the nonradiative lifetime even more, due to the relocalization of excitons. Therefore, at $T= 4$~K, the radiative efficiency is close to $1$ for low exciton densities and the effective lifetime equals the radiative lifetime. At higher temperature the nonradiative time is less affected by the exciton density, due to the more important contribution of the thermal energy to the exciton delocalization.
 Even at low density the recombination time never reaches the radiative limit (which is temperature independent). This leads to the onset of the nonradiative horizon, well reproduced by this model.
Finally, while both $D$ and $\tau$ change significantly with $n$, their product $D\tau$ changes much less. That is why the simple model of constant diffusion coefficient and constant recombination time presented in Fig.~\ref{fig:sampleA}(a) could successfully fit at least the PL blue shift.

\section{Conclusions}

\label{sec:end}

In conclusion, we have studied GaN/(Al,Ga)N $c$-plane quantum well structures
using space and time-resolved PL spectroscopy.
%
%
In the emission pattern that we represent in the distance/energy
coordinate space two distinct features are identified:
(1) an arrow-shaped pattern up to $20$ $\mu$m from the excitation spot,
which does not depend on the temperature (up to $80$~K)  and (2) a long
range tail of the emission at constant energy, which disappears when
the temperature is increased.
The combination of spectral, spatial and temporal imaging of the
exciton emission, and a detailed model of the transport and
recombination processes, allow to determine the most relevant
phenomena explaining the short- and long-distance limits, and the
short and long timescales.
%
%
%
%
The arrow-shaped pattern  is due to the density-activated exciton
diffusion away from the excitation spot, and the density dependent
emission energy.
%
%
In contrast with GaAs-based structures, exciton transport is
dominated by density-activated diffusion, rather than by the drift
current.
The long-range PL tail is attributed to secondary excitons, directly
created at long distances by higher-energy photons emitted around $r
= 0$ by the QW itself and guided along the sample plane.
This emission is efficiently quenched at high temperature beyond a
certain distance, that we identify as the nonradiative horizon.
\cite{Grosso2009}
%

The results obtained in this work suggest that GaN/AlGaN QWs constitute
a very promising system for the study of dipolar excitons and exciton-based devices.
 We have demonstrated that exciton transport takes place over more than 10 $\mu$m
 up to $80$ K. Reducing the dislocation density by using GaN rather than
 sapphire substrate could substantially increase this distance.
  Consequently, such structures are promising for the studies of cold
  exciton gases and related quantum phenomena.
  The possibility to spatially separate the generation spot from the
  cold excitons reservoir could be crucial to achieve the quantum regime.
  %
  %
  In terms of the  transport mechanisms, it could be interesting
  to develop new models, taking into account energy relaxation and the
  formation of the phonon replica.
  The latter phenomenon is quite strong in the  wurtzite semiconductors,
  compared to their cubic counterparts, and might affect the propagation of excitons.
  Other polar wide-band gap materials, such as ZnO/ZnMgO with excitons
  stable at room temperature  is another possible research direction.

\section{Acknowledgements}
We are grateful to K.~V.~Kavokin, M. I. Dyakonov, L. V. Butov and D.
Scalbert for valuable discussions and acknowledge the support of EU
ITN INDEX PITN-GA-2011-289968.

\bibliographystyle{apsrev4-1}
\bibliography{biblio}


\end{document}